\documentclass[prl,twocolumn,showpacs,superscriptaddress,floatfix]{revtex4}
\usepackage{graphicx}
\usepackage{dcolumn}

\newcommand{\vdc}{\ensuremath{V_{\text{dc}}}}
\newcommand{\vac}{\ensuremath{v_{\text{ac}}}}
\newcommand{\vrf}{\ensuremath{V_{\text{rf}}}}
\newcommand{\vrfq}{\ensuremath{v_{\text{rf}}^{\text{QPC}}}}

\newcommand{\vm}{\ensuremath{v_{m}}}
\newcommand{\fo}{\ensuremath{f_{0}}}

\newcommand{\dq}{\ensuremath{\delta q}}

\newcommand{\ehz}{\ensuremath{e/\sqrt{\mathrm{Hz}}}}
\newcommand{\aehz}[1]{\ensuremath{#1\:\ehz}}
\newcommand{\e}[1]{\ensuremath{\times 10^{#1}}}
\newcommand{\units}[1]{\ensuremath{\mathrm{#1}}}
\newcommand{\amount}[2]{\ensuremath{#1\:\units{#2}}}
\newcommand{\cp}{\ensuremath{C_{p}}} 

\newcommand{\vg}{\ensuremath{V_{g}}}

\newcommand{\sis}{\ensuremath{S_{I}}}
\newcommand{\gqpc}{\ensuremath{G_{\text{QPC}}}}
\newcommand{\go}{\ensuremath{G_{0}}}
\newcommand{\zo}{\ensuremath{Z_{0}}}
\newcommand{\wo}{\ensuremath{\omega_{0}}}
\newcommand{\tn}{\ensuremath{T_{n}}}

\newcommand{\algas}{GaAs/AlGaAs}

\newcommand{\pn}{\ensuremath{P_{n}}}
\newcommand{\pin}{\ensuremath{P_{\text{in}}}}
\newcommand{\ps}{\ensuremath{P_{s}}}
\newcommand{\pnt}{\ensuremath{P_{n}^{T}}}
\newcommand{\pna}{\ensuremath{P_{n}^{A}}}
\newcommand{\pne}{\ensuremath{P_{n}^{E}}}
\newcommand{\pnei}{\ensuremath{\mathcal{P}_{n}^{E}}}
\newcommand{\vs}{\ensuremath{v_{s}}}

\begin{document}

\title{Shot-Noise-Limited Operation of a Fast Quantum-Point-Contact Charge Sensor}

\author{Madhu Thalakulam}
\affiliation{Department of Physics and Astronomy, Rice University, Houston, TX 77005 USA}
\affiliation{Rice Quantum Institute, Rice University, Houston, TX 77005 USA}

\author{W. W. Xue}
\affiliation{6127 Wilder Laboratory, Department of Physics and Astronomy,
Dartmouth College, Hanover, NH, 03755 USA}

\author{Feng Pan}
\affiliation{6127 Wilder Laboratory, Department of Physics and Astronomy,
Dartmouth College, Hanover, NH, 03755 USA}

\author{Z. Ji}
\affiliation{Department of Physics and Astronomy, Rice University, Houston, TX 77005 USA}

\author{J. Stettenheim}
\affiliation{6127 Wilder Laboratory, Department of Physics and Astronomy, Dartmouth College, Hanover, NH, 03755 USA}

\author{Loren Pfeiffer}
\affiliation{Bell Laboratories, Lucent Technologies, Inc., Murray Hill, New Jersey 07974}

\author{K. W. West}
\affiliation{Bell Laboratories, Lucent Technologies, Inc., Murray Hill, New Jersey 07974}

\author{A. J. Rimberg}\email{ajrimberg@dartmouth.edu}
\affiliation{6127 Wilder Laboratory, Department of Physics and Astronomy,
Dartmouth College, Hanover, NH, 03755 USA}

\begin{abstract}
We have operated a quantum point contact (QPC) charge detector in
a radio frequency (RF) mode that allows fast charge detection in
a bandwidth of tens of megahertz.  We find that the charge
sensitivity of the RF-QPC is limited not by the noise of a
secondary amplifier, but by non-equilibrium noise \sis\ of the
QPC itself.  We have performed frequency-resolved measurements of
the noise within a \amount{10}{MHz} bandwidth around our carrier
wave.  When averaged over our bandwidth, we find that \sis\ is in good agreement with the theory of photon-assisted shot noise.  Our measurements also reveal strong frequency dependence of the noise, asymmetry with respect to the carrier wave, the appearance of sharp local maxima that are correlated with
mechanical degrees of freedom in the sample, and noise suppression indicative of many-body physics near the 0.7 structure.  
\end{abstract}


\maketitle

All measurements, including electrical amplification, are subject to quantum mechanical limits \cite{Caves:1982,Devoret:2000}: a standard measurement of a quantum system must add noise with a strictly determined minimal size.  To reach this quantum limit, the output noise of a measurement system must be dominated by the intrinsic noise of an initial quantum amplifier and not that of a subsequent classical one \cite{Clerk:2004b,Korotkov:2003}.   This requires that shot noise  arising from the flow of current through the quantum amplifier dominates the measurement system noise.  Here, we report shot-noise limited operation of a quantum-point-contact (QPC) charge sensor in a radio-frequency (RF) mode analogous to that used for single electron transistors \cite{Schoelkopf:1998}.

QPCs, one of the simplest nanoscale systems, are surprisingly complex.  Recently, study of QPC has been focused on two areas in particular.  First, there is a strong interaction between electronic and mechanical degrees of freedom in GaAs-based QPCs, allowing both detection of mechanical resonances using a QPC as a detector \cite{Cleland:2002} and synchronized transport of electrons through QPCs in the tunneling regime \cite{Shilton:1996}.  Second, there is both experimental \cite{Expt07,Cronenwett:2002} and theoretical \cite{Theory07} evidence of the formation of a many-body magnetic impurity state in QPCs that manifests itself as an anomalous plateau in the QPC conductance at $\gqpc\approx 0.7\go$ where $\go=2e^{2}/h$.   

In this Letter, we use frequency-resolved measurements of shot noise \cite{Shotnoise,Shotnoise07} in a heretofore unexplored limit to characterize our RF-QPCs\@.  We find the shot noise in the vicinity of the carrier wave frequency \fo\ shows surprising frequency dependence and reflects both the physics of the 0.7 structure and the interplay between vibrational and electronic degrees of freedom.  Coupling of electronic and mechanical degrees of freedom and the  presence of a local moment in a QPC do not appear to have been considered previously with regard to its potential as a quantum limited charge detector.  Our measurements of the intrinsic noise and charge sensitivity of an RF-QPC charge detector lie at the intersection of these three areas of investigation.  
\begin{figure}[ht]
\begin{center}
\includegraphics[width=7.5cm]{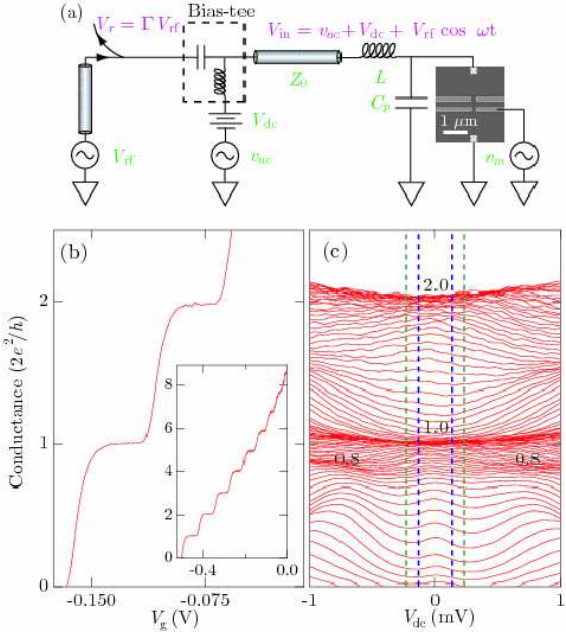} 
\caption{\label{fig1} (Color online)  (a) Schematic diagram of the measurement circuit.  The electron micrograph shows the QPC geometry; only one constriction is used in the measurements.  The RF carrier wave is applied via a directional coupler, which also directs the reflected wave to a cryogenic HEMT amplifier with noise temperature \amount{2.3}{K} followed by a GaAs FET amplifier at room temperature.   There was  \amount{48}{dB} of attenuation in the input RF lines. All dc lines passed through cascaded $\pi$-type, $RC$ and microwave filters. A circulator between the tank circuit and the HEMT amplifier isolated the sample from noise sources on the output line. (b) \gqpc\  versus gate voltage \vg\ at zero magnetic field and $T=\amount{25}{mK}$.   Inset: \gqpc\  after exposure of the sample to light, showing multiple conductance plateaus.  (c) Nonlinear conductance $\gqpc(\vdc)$.  Measurements were performed for a series of values of \vg\ with spacing $\Delta\vg=\amount{1}{mV}$  and plotted without offset.  The vertical dashed lines indicate the estimated rms rf voltage applied to the QPC for subsequent noise measurements.  }
\end{center}
\end{figure}

Our QPCs were formed via the split gate technique in a \algas\
heterostructure containing a 2DEG with sheet density
$n_{s}=\amount{1.3\e{11}}{cm^{-2}}$ and mobility
$\mu=\amount{7.4\e{6}}{cm^{2}V^{-1}s^{-1}}$  located \amount{100}{nm}
beneath the heterostructure surface.  We fabricated two samples, A and
B\@.  Except where noted all data shown is from sample A; results for
sample B were similar. Measurements were performed in a dilution
refrigerator with a base temperature of $T=\amount{25}{mK}$ and
effective electron temperature $T_{e}\approx\amount{80}{mK}$ .  The QPCs
were imbedded in an $LC$ tank circuit consisting of a Nb spiral chip
inductor with  $L=\amount{140 (125)}{nH}$ for sample A (B), parasitic
capacitance  $\cp=\amount{0.28 (0.25)}{pF}$ and resonant frequency
$\fo=1/2 \pi\sqrt{LC}=\amount{800 (900)}{MHz}$.  A
bias-tee in our rf circuitry [Fig.~\ref{fig1}(a)] allowed application of
an RF (\vrf) signal for microwave
reflectometry measurement of the QPC charge sensitivity and noise
\cite{Schoelkopf:1998,Lu:2003} and near-dc voltages (\vac\ and \vdc) for lockin measurements of the QPC conductance \gqpc.  Conductance data for our QPCs
($\vac=\amount{20}{\mu V}$ rms at \amount{13}{Hz}) show well-defined
plateaus in \gqpc\ versus the voltage \vg\ applied to the split gates
[Fig.~\ref{fig1}(b)].

Application of  a dc voltage  \vdc\ allowed measurement of nonlinear differential conductance  $\gqpc(\vdc)$ versus both \vg\ and \vdc.  For $T<\amount{500}{mK}$, we observed a peak in  \gqpc\ around  $\vdc=0$ for QPC conductance in the range $0<\gqpc<\go$ [Fig.~\ref{fig1}(c)].  This zero-bias anomaly (ZBA) has been studied previously \cite{Cronenwett:2002} and interpreted as an indication of the onset of Kondo physics in the QPC \cite{Theory07},  as has an additional plateau at finite bias ($\vdc\approx\amount{700}{\mu V}$) for which $\gqpc\approx 0.8\go$ \cite{Cronenwett:2002}.   These measurements of  $\gqpc(\vdc)$ provide clear evidence that the physics associated with the 0.7 structure is present in our QPCs and are indicative of their high quality.  

\begin{figure}[h]
\begin{center}
\includegraphics[width=7.5cm]{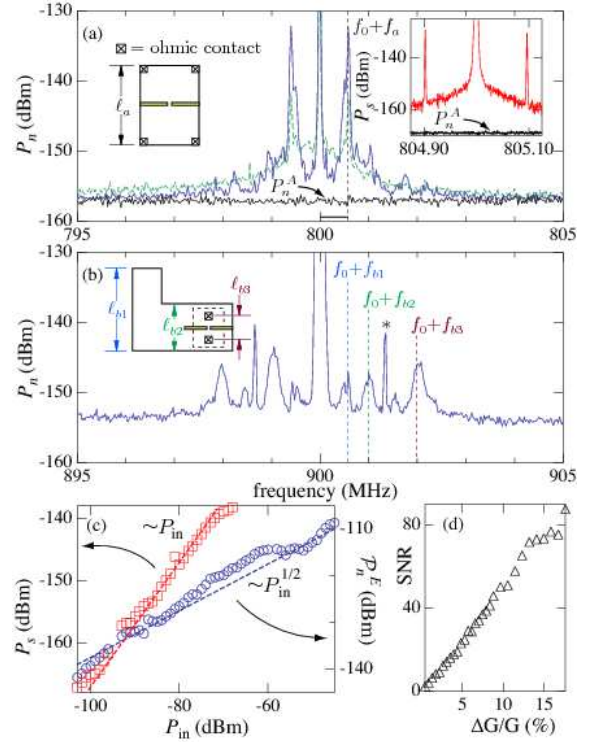} 
\caption{\label{fig2} (Color online) (a) Output power spectrum \pn\ of an RF-QPC (sample A) for $\pin=\amount{-98}{dBm}$ (dashed green) and \amount{-88}{dBm} (solid blue) and HEMT noise floor \pna.  Both \pin\ and \pn\ are referred to the input of the HEMT amplifier.  Right inset: output  of the RF-QPC subject to conductance modulation and HEMT noise floor \pna.  Measurement is for $\pin=\amount{-57}{dBm}$  and $\gqpc\approx 0.5\go$.  Left inset: sample A geometry. (b) \pn\ for sample B.  Inset: sample B geometry.  (c)  \ps\ (squares) and \pnei\  (circles) versus \pin.  Dashed lines are guides to the eye that scale as \pin\  and $\pin^{1/2}$.   (d) Signal-to-noise ratio for the RF-QPC on a linear scale versus conductance modulation.}
\end{center}
\end{figure}

To operate our QPC as a charge detector, we tuned \vg\ to maximize  $d\gqpc/d\vg$  (typically $\gqpc\approx 0.5\go$) at $\vdc=0$ and applied an rf carrier wave $\vrf\cos\wo t$ where $\wo=2\pi\fo$ to the tank circuit.  Some portion  $\Gamma\vrf\cos\wo t$  of the wave is reflected (the reflection coefficient  $\Gamma$ of the tank circuit depends on \gqpc) and is measured at the output of our amplifier chain \cite{Schoelkopf:1998,Lu:2003}. The RF-QPC bandwidth is determined by the width  $\Delta f =\fo/Q\approx\amount{60}{MHz}$ of the tank circuit resonance, allowing very fast charge detection.  For a QPC coupled to a quantum dot, an electron tunneling event typically changes \gqpc\ by 1--3\% \cite{Vandersypen:2004}.  To mimic this effect, we apply a small ac voltage  \vm\ [Fig.~\ref{fig1}(a)] at \amount{97}{kHz} to one QPC gate so that  $\Delta\gqpc/\gqpc\approx 2.7\%$.  The RF-QPC output shows side peaks at $\fo\pm\amount{97}{kHz}$  indicative of amplitude modulation riding on a broad noise background [right inset, Fig.~\ref{fig2}(a)].  We estimate the charge sensitivity of the QPC to be  $\dq\approx\aehz{5\e{-4}}$ referred to a hypothetical quantum dot.

Two aspects of the noise limiting the QPC sensitivity are striking: first, it is larger than the noise \pna\ of the HEMT amplifier, which usually limits the performance of RF-SETs; second, it is frequency-dependent rather than white.  To investigate, we measure the spectrum of reflected noise power \pn\ in a \amount{10}{MHz} bandwidth around \fo\ for different values of the input power  \pin\ and with no conductance modulation  [Fig.~\ref{fig2}(a)].  In addition to broadband noise that decreases away from \fo, there are large peaks in \pn at  $\fo\pm\amount{580}{kHz}$.  For $\pin=\amount{-98}{dBm}$ the broadband noise is clearly visible and the peaks are relatively small; for larger input power  $\pin=\amount{-88}{dBm}$ the broadband noise decreases while the peaks at  $\fo\pm\amount{580}{kHz}$ become more pronounced.  \pn\ for sample B shows similar peaks but a more complex spectrum.  

Since the measured noise  \pn\ depends on \pin\  (and on  \gqpc, see below), it is associated with the sample.  There are two broad categories into which such noise might fall: modulation noise, for which the current through the QPC is amplitude modulated; and shot noise \cite{Kogan:1996}.  Modulation noise scales with input power as $\pn\propto\pin$  whatever its origin, whether motion of trapped charges in the substrate, electromagnetic noise coupled to the QPC gates, mixing due to the QPC nonlinearity, or some other source.  Shot noise, in contrast, scales as $\pin^{1/2}$.  

In our experiment shot noise arises from the partition noise of electron-hole pairs created by the RF voltage \vrfq\ across the QPC \cite{Reydellet:2003}; for an ideal matching network this (rms) voltage is given by  $\vrfq=2Q\vrf=2Q\sqrt{\pin\zo}$.  Such ``photon assisted'' shot noise (PASN) has been examined theoretically \cite{PASNTheory} and measured both in normal metals \cite{Schoelkopf:1998a} and QPCs \cite{Reydellet:2003}.  Previous work has studied  PASN at a frequency $\omega$ much less than the drive frequency \wo.  Here, we measured PASN for $\omega\approx\wo$.  Assuming energy-independent transmission coefficients \tn\ it can be shown that the spectral density of photon-assisted shot noise is given by $\sis (\omega ,\wo ) = \frac{{4e^2 }}{h}\sum\limits_n {\tn (1 - \tn)} \sum\limits_{l =  - \infty }^\infty  {(\hbar \omega  + l\hbar \wo )\,J_l^2 (\alpha )\coth \left[ {\frac{{\hbar \omega  + l\hbar \wo }}{{2k_B T}}} \right]}$ where $\alpha=\sqrt{2}e\vrfq/\hbar\wo$.  For low temperature and $\alpha\gg 1$ the infinite sum can be evaluated easily and scales as $\alpha\propto\pin^{1/2}$.  In addition to shot noise, \pn\ includes contributions \pnt\  from thermal noise and  \pna\ from the HEMT amplifier that we account for by extracting the excess noise $\pne=\pn-\pnt-\pna$  from our raw data.  The prediction for $\sis(\omega,\wo)$  above allows us to determine the origin of the excess noise  \pne\ by measuring its dependence on  \pin\ and $\gqpc=\go\sum_{n}\tn$.

We varied \pin\  over a six decade range, and measured both the power  \ps\ in a charge modulation signal and the integrated excess noise $\pnei=\int\pne df$ in a 4.8 MHz bandwidth above \fo\ (with no charge modulation).  We find $\ps\propto\pin$  over a range of three decades in  \pin\ before the RF-QPC response begins to saturate [Fig.~\ref{fig2}(c)].  The linearity of the RF-QPC in this range is excellent: the SNR for the modulation signal rises linearly with increasing $\Delta\gqpc/\gqpc$ up to   $\Delta\gqpc/\gqpc=15$\%  [Fig.~\ref{fig2}(d)].  In contrast \pnei\  scales as $\pin^{1/2}$ over a nearly five decade range, eliminating modulation noise as the source of \pne.  

\begin{figure}[h]
\begin{center}
\includegraphics[width=7.5cm]{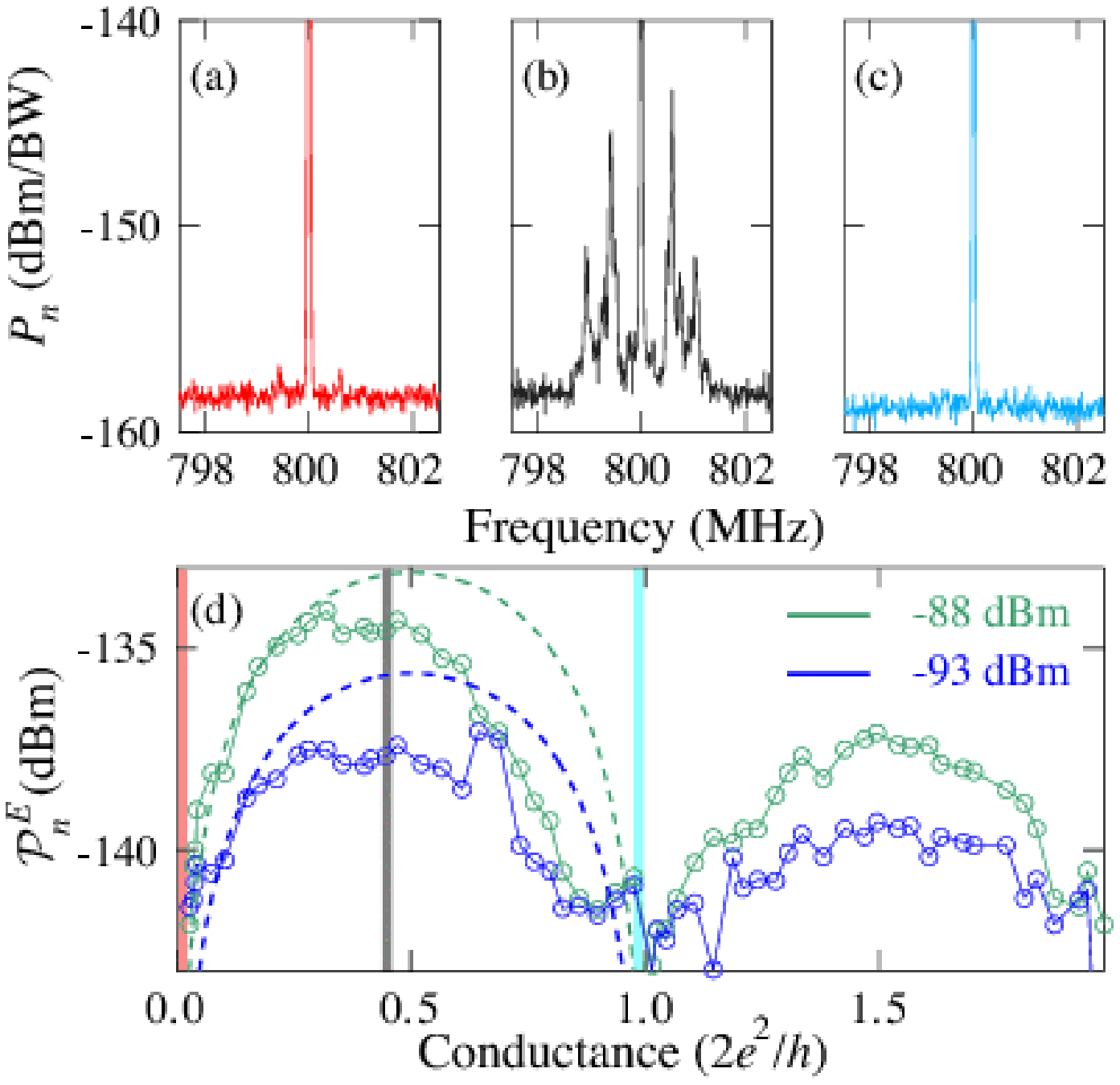} 
\caption{\label{fig3} (Color online) (a)--(c) \pn\ for $\pin=\amount{-88}{dBm}$   and  $\gqpc\approx 0$  (a),  $0.5\go$ (b), and \go\  (c).  (d) \pnei\ versus \gqpc\  for $\pin=\amount{-88}{dBm}$ (top) and\amount{-93}{dBm} (bottom), corresponding to $\vrf^{\text{QPC}}=\amount{230}{\mu V}$ and \amount{130}{\mu V} respectively, as indicated relative to the QPC $IV$ characteristics in Fig.~\ref{fig1}(c) by the vertical dashed lines.  There is no rise in the noise floor for  $\gqpc\approx\go$ versus  $\gqpc\approx 0$ [compare (c) and (a)], indicating that there is no significant sample heating for these input powers.  To compare with theory, we integrated the predicted PASN power over the same bandwidth as for \pnei\ and converted to noise power at the HEMT amplifier (dashed lines).  Current noise $\sis(\omega,\wo)$ in the QPC is transformed by an ideal $LC$ matching network into noise power $(2L/\cp\zo)\sis(\omega,\wo)$  at the input to the HEMT amplifier, where $\zo=\amount{50}{\Omega}$  is the impedance of the coaxial cable connecting it to the tank circuit. The results were shifted downward by \amount{3.9}{dB} but no other fitting parameter was used.   The reduction of the measured noise relative to theory is likely due to losses in the matching network.  }
\end{center}
\end{figure}

We also measured \pnei\  versus  \gqpc\ over the range $0<\gqpc<2\go$  for two different values of  \pin, corresponding to \vrfq\ indicated by the dashed lines in Fig.~\ref{fig1}(c).  \pne\ vanishes for $\gqpc\approx 0$, is maximal for  $\gqpc\approx 0.5\go$, and vanishes again for    $\gqpc=1.0\go$ [Fig.~\ref{fig3}(a)--(c)].   A more detailed set of measurements [Fig.~\ref{fig3}(d)] confirms that the magnitude of \pnei\ is well described by the shot noise $\sis(\omega,\wo)$  integrated over the same bandwidth and converted to voltage noise \cite{Korotkov:1999} by the tank circuit [Fig.~\ref{fig3}(d), dashed lines].   Interestingly, \pnei\ is noticeably suppressed for  \gqpc\ in the vicinity of $0.7\go$, in agreement with recent measurements of dc shot noise in QPCs \cite{Shotnoise07}.  These observations, combined with the scaling as $\pin^{1/2}$  described earlier, conclusively identify shot noise as the source of the excess noise \pne. 

In contrast to the calculated  $\sis(\omega,\wo)$,  \pne\ depends strongly on $\omega$. It is not uncommon, however, for noise to show spectral features corresponding to physical excitations of a system \cite{Balatsky:2006}.  We hypothesize that a surface acoustic wave (SAW) with a half-wavelength equal to a typical sample dimension $\ell$ is excited in the piezoelectric GaAs substrate by the rf drive and take the SAW frequency to be $f_{\ell}=\beta \vs/2\ell$ where $\vs=\amount{3010}{m/s}$ is the speed of sound in GaAs and $\beta=1.05$ is a scaling parameter.  We expect the SAW to produce features in \pn\ at $\fo\pm f_{\ell}$.  For sample A there is only one relevant length scale $\ell_{a}$ [left inset, Fig.~\ref{fig2}(a)] for which $f_{a}\approx\amount{580}{kHz}$.  Agreement of $f_{a}$ with the offset of the noise peaks in Fig.~\ref{fig2}(a) from \fo\ is remarkable.  For sample B there are three relevant length scales [inset, Fig.~\ref{fig2}(b)], $\ell_{b1}$ (\amount{2.8}{mm}), $\ell_{b2}$ (\amount{1.6}{mm}) and $\ell_{b3}$ (\amount{0.8}{mm}) with corresponding frequencies $f_{b1}$ (\amount{560}{kHz}), $f_{b2}$ (\amount{990}{kHz}) and $f_{b3}$ (\amount{1.98}{MHz}).  For each $f_{bi}$ there is a broad peak in \pn\ at $\fo\pm f_{bi}$ that scales as $\pin^{1/2}$, providing strong evidence of coupling between shot noise and mechanical degrees of freedom in our RF-QPCs.  The peak marked by the asterisk scales as \pin\, identifying it as modulation noise.

\begin{figure}[h]
\begin{center}
\includegraphics[width=7.5cm]{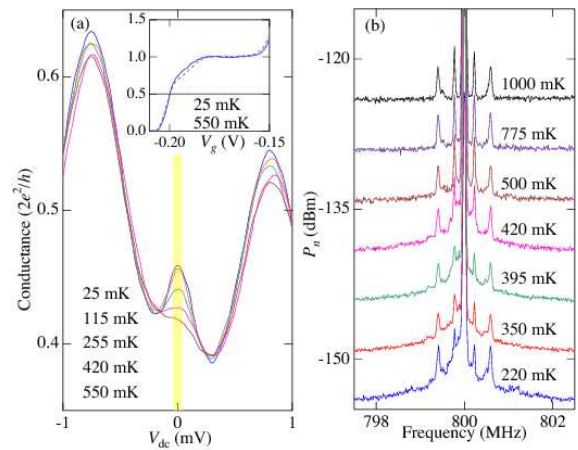} 
\caption{\label{fig4} (Color online)  (a) $\gqpc(\vdc)$ near $0.5\go$ versus \vdc\ for  $T=25$ to \amount{550}{m K}.   The size of the rms rf bias $\vrf^{\text{QPC}}=\amount{40}{\mu V}$ for the noise measurements in (b) is indicated by the yellow shaded region around $\vdc=0$.  Inset: differential conductance versus gate voltage for $T=25$ (solid) and \amount{550}{mK} (dashed).  (b) Noise power \pn\  frequency for $T = 225$ to \amount{1000}{mK}, near $\gqpc\approx 0.5\go$  and for $\pin=\amount{-103}{dBm}$.  Successive curves are offset by \amount{5}{dB} for clarity.  Peaks at  $\fo=\pm\amount{250}{kHz}$ scale as  \pin\ and are due to modulation noise. }
\end{center}
\end{figure}

For \pin\ used in Fig.~\ref{fig3}  \vrfq\ was sufficiently large to drive the QPC away from the ZBA\@.  However, we were able to measure \pn\ near $\gqpc\approx 0.5\go$ for $\pin=\amount{-103}{dBm}$  for which \vrfq\ lies entirely within the ZBA at a series of temperatures [Fig.~\ref{fig4}].  Interestingly, the broadband noise is noticeably asymmetric with respect to \fo\ [Fig.~\ref{fig4}(b)] for $T<\amount{500}{mK}$.  As the temperature is raised, the broadband noise both weakens and becomes more symmetric, so that for $T>\amount{500}{mK}$ it has nearly vanished.  In contrast, the peaks at  $\fo\pm\amount{580}{kHz}$ are clearly visible for $T=\amount{1}{K}$, suggesting different physical origins for the two phenomena. Note that the ZBA has a temperature dependence similar to that of the broadband noise, weakening rapidly for temperatures above \amount{115}{mK} and nearly vanishing for  $T>\amount{550}{mK}$.  The asymmetry in \pn\  also vanishes when \vrfq\ is far out of the ZBA: in Fig.~\ref{fig2}(a) some asymmetry is visible in \pn\  for $\pin=\amount{-98}{dBm}$ but not for $\pin=\amount{-88}{dBm}$.  Similar dependence on $T$ and \pin\ for the broadband noise and the ZBA suggest they may be related; further experiments are needed for a conclusive demonstration.  

In conclusion, we have operated an RF-QPC at the shot noise limit.   The noise both shows coupling to mechanical degrees of freedom in the sample and reflects the many-body physics of the 0.7 structure.  Our results suggest that such phenomena may have important implications for the ultimate charge sensitivity of the QPC and how nearly it can approach the quantum limit.  Our results have immediate implications for study of spin-based quantum information processing in quantum dots \cite{Petta:2005,Koppens:2006}.  The techniques employed here may also be applicable to studies of noise in other semiconductor devices such as quantum dots in the Kondo regime.

This work was supported by the NSF under Grant No.\ DMR-0454914, by the ARO under Agreement No.\ W911NF-06-1-0312 and by the NSA, LPS and ARO under Agreement No. W911NF-04-1-0389.  We thank M. Blencowe for many helpful conversations.  


\end{document}